\documentclass[sn-mathphys,Numbered]{sn-jnl}
\usepackage{graphicx}%
\usepackage{multirow}%
\usepackage{amsmath,amssymb,amsfonts}%
\usepackage{amsthm}%
\usepackage{mathrsfs}%
\usepackage[title]{appendix}%
\usepackage{xcolor}%
\usepackage{textcomp}%
\usepackage{manyfoot}%
\usepackage{booktabs}%
\usepackage{algorithm}%
\usepackage{algorithmicx}%
\usepackage{algpseudocode}%
\usepackage{listings}%
\usepackage{latexsym}
\usepackage{bm}
\usepackage[caption=false,lofdepth,lotdepth]{subfig}
\usepackage{soul}
\usepackage[normalem]{ulem}
\usepackage{hyperref} 
\hypersetup{%
	pdfpagemode=FullScreen,
	pdfstartpage=1,
	pdfmenubar=true,
	pdftoolbar=true,
	colorlinks = true,
	linkcolor=blue,
	citecolor=blue,
	urlcolor=blue,
	bookmarksopen=false
}

\begin{document}

\title{Relative dynamics of quantum vortices and massive cores in binary BECs}

\author[1]{\fnm{Alice} \sur{Bellettini}}\email{alice.bellettini@polito.it}

\author*[2,3]{\fnm{Andrea} \sur{Richaud}}\email{andrea.richaud@upc.edu}

\author[1]{\fnm{Vittorio} \sur{Penna}}\email{vittorio.penna@polito.it}

\affil*[1]{\orgdiv{Department of Applied Science and Technology}, \orgname{Politecnico di Torino}, \orgaddress{\street{Corso Duca degli Abruzzi, 24}, \city{Torino}, \postcode{10129}, \country{Italy}}}

\affil[2]{\orgname{Scuola Internazionale Superiore di Studi Avanzati (SISSA)}, \orgaddress{\street{Via Bonomea 265}, \city{Trieste}, \postcode{34136}, \country{Italy}}}

\affil[3]{\orgdiv{Departament de F\'isica}, \orgname{Universitat Polit\`ecnica de Catalunya}, \orgaddress{\street{Campus Nord B4-B5}, \city{Barcelona}, \postcode{08034}, \country{Spain}}}

\abstract{We study the motion of superfluid vortices with filled massive cores. Previous point-vortex models already pointed out the impact of the core mass on the vortex dynamical properties, but relied on an assumption that is questionable in many physical systems where the immiscibility condition is barely satisfied: the fact that the massive core always lays at the very bottom of the effective confining potential constituted by the hosting vortex. Here, we relax this assumption and present a new point-vortex model where quantum vortices are harmonically coupled to their massive cores. We thoroughly explore the new dynamical regimes offered by this improved model; we then show that the functional dependence of the system normal modes on the microscopic parameters can be correctly interpreted only within this new generalized framework. Our predictions are benchmarked against the numerical simulations of coupled Gross-Pitaevskii equations for a realistic mixture of atomic Bose-Einstein condensates.}


\maketitle

\section{Introduction}

Boson mixtures of dilute gases formed by two components can feature vortex configurations with massive cores, whose rich phenomenology has attracted considerable interest in the last two decades. Controlled by the interplay of repulsive inter-species and intra-species interactions, one of the two components can thicken in the vortex cores of the other component, where the vortices play the role of a local trapping potential. This leads to a complete phase separation, the immiscibility regime, for sufficiently strong inter-species interaction. 

Vortices characterized by filled cores, experimentally realized in mixtures with different hyperfine components \cite{Matthews1999,Anderson2000}, have been theoretically investigated in Refs. \cite{Ho1996,Garcia2000,Skryabin2000,Mcgee2001} to explore their stability and phase-separation properties, and have been involved as well in the study of multi-vortex lattices in rotating condensates. Here phase separation plays a crucial role in the formation of such structures \cite{Mueller2002}, in determining the properties of their phase diagram \cite{Kasamatsu2003}, and in the mixing dynamics of Skyrmion lattices \cite{Schwe2004}. 
In the last decade, a renewed interest for this class of systems has highlighted unexplored dynamical behaviors and complex properties, due to the possibility of combining analytical tools with improved numerical techniques and resources. A variety of aspects and effects have been investigate which include, in the immiscible regime, the formation of vortices with massive cores in the form of stable vortex–bright-soliton structures \cite{Law2010}, vortex-bright soliton dipoles and the tunnelling of the soliton component \cite{Pola2012}, and their dynamics with unequal dispersion coefficients \cite{Charalampidis2016}.
On the other hand, in the miscible regime, the two species do not separate and exhibit the formation of vortex states representing magnetic defects \cite{Gallemi2018}, and configurations with vortices \cite{Kuopanportti2019,Han2022} or half-vortices \cite{Eto2011} in both the components, where the cores correspond to density peaks of the other species. Again in the miscible regime, the phase diagram of complex topological-defect states, for a rotating binary condensate, has been determined as a function of the spatial separation and the angular velocity \cite{Mason2011}, while the angular-momentum exchange in vortex-bright soliton structures has been related to the miscibility of the two components \cite{Mukherjee2020}.

Recently, it has been shown that the dynamics of 2D massive vortices in a mixture of components $a$ and $b$ is described by Lorentz-like electromagnetic ODEs \cite{Richaud2020,Richaud2021}, where the $a$ density plays the role of the magnetic field intensity and the vortex interaction force is analogous to the two-dimensional (2D) electric force. The $b$ core mass adds a classical kinetic energy term. This point-like model allows for the description, in the immiscible regime, of a system of stable vortices via a set of coordinates or degrees of freedom, where the comparison with the mean-field Gross-Pitaevskii (GP) equations has been satisfactory. 
However, one of the assumptions of the aforementioned point-vortex model (see also Refs. \cite{Richaud2021,Richaud2022r_k,Caldara2023}) is that the center of the massive core \emph{always} coincides with that of the hosting quantum vortex. 
This is, a priori, questionable, especially when the immiscibility condition is barely met, as the massive core may, in principle, oscillate around the vortex center.

Here, we overcome this  constraint and extend the massive point-vortex model by treating the vortex and its massive core as two different dynamical objects, associated to two different sets of dynamical variables.
In the point-like derivation, we assimilate the coupling energy to that of a restoring spring. Besides, we extract numerical data from the simulation of GP equations and use them to investigate the vortex-massive core \emph{real} dynamics and to validate our analytical model. As previously observed \cite{Richaud2021}, the first striking effect of the second species is a qualitative change of trajectory. Normally, a massless vortex in a 2D disc moves of uniform circular motion. In presence of a second species, characteristic radial oscillations arise as the hallmark of the inertial character of the core mass.
Specifically, we find that our new model overcomes the previous one in predicting the dependency of the radial oscillations from the inter-species coupling parameter $g_{ab}$. This dependency was not intrinsic in the previous model as it did not include the parameter $g_{ab}$ at all. We wish to remark that the predictions of our analytical model are confirmed by the analysis of the system eigenfrequencies as extracted from GP simulations, despite the fact that 
there is no significant relative motion between the vortex and its massive core,
in agreement with the recent results of Ref. \cite{Doran2022}.

Our manuscript is organized as follows: in Sec. \ref{sec:Model} we present the physical system and derive an effective point-vortex model where vortices are harmonically coupled to massive cores. In Sec. \ref{sec:Normal_modes}, we extensively explore the new dynamical regimes and the effects predicted by such improved point-vortex model, pointing out some important fully-analytical properties. Section \ref{sec:GP_results} is concerned with numerical simulations of coupled GP equations, followed by suitable extractions of the vortices' and massive cores' trajectories. The comparison between numerical results and analytical predictions is extensively discussed. Eventually, Sec. \ref{sec:Conclusions} is devoted to the concluding remarks and sketches some possible future research directions.


\section{Point-like model for vortices harmonically coupled to massive cores}
\label{sec:Model}
We consider a two-component BEC strongly confined along the $z$-direction by a harmonic trap characterized by a harmonic-oscillator length $d_z$. Its dynamics is described, at the mean-field level, by two coupled GPEs
\begin{equation}
\label{eq:H_GPE}
i\hbar\frac{\partial \psi_i}{\partial t} =\! \!\left(\!-\frac{\hbar^2\nabla^2}{2m_i} + V_{\rm tr} 
+ \sum_{j=a,b}g_{ij}|\psi_j|^2\!\right)\psi_i, \; i=a,b
\end{equation}
where $\psi_a$ and $\psi_b$ are the macroscopic wavefunctions associated to the two components, whose norm is $N_i=\int|\psi_i|^2\,\mathrm{d}^2 r$, with $i=a,\,b$. The model parameters $m_a$ and $m_b$ represent the atomic masses of the two components, while $g_{ij}=\sqrt{2\pi}\hbar^2 a_{ij}/(m_{ij} d_z)$ represent the intra- and inter-species couplings,
which depend on the s-wave scattering lengths $a_{ij}$ and on the effective masses
$m_{ij}=(m_{i}^{-1}+m_{j}^{-1})^{-1}$ \cite{Hadzibabic2011}.
The term $V_{\rm tr}$ is the confining potential, which is taken to be a circular hard-wall potential; 
in the strongly-interacting Thomas-Fermi regime ($N_i a_{ii}\gg d_z$), the equilibrium density is almost homogeneous within such a trap and thus better represents a portion of an extended quasi-2D superfluid system.  Equations (\ref{eq:H_GPE}) in the immiscible regime $g_{ab}>\sqrt{g_a g_b}$ admit a notable class of solutions, namely a vortex in component $a$ filled by $N_b\ll N_a$ $b$-component particles. 

\subsection{Lagrangian variational approach}

In general, the essential time evolution of the complex fields $\psi_a$ and $\psi_b$  can be captured according to a time-dependent variational approximation \cite{PerezGarcia1996,Kim2004,Richaud2021}. This technique allows one to bypass the need of solving Eqs. (\ref{eq:H_GPE}), and to reduce the dynamics of complex fields $\psi_a$ and $\psi_b$ to that of few selected time-dependent variational parameters. For the present problem, the ansatz
\begin{equation}
    \label{eq:psi_a}
    \psi_a=\sqrt{n_a-n_a e^{-|\bm{r}-\bm{r}_v(t)|^2/\sigma_a^2 }
    }\,e^{i\theta_a} \,\quad \text{for} |\bm{r}|<R
\end{equation}
is such that its density field $|\psi_a|^2$ is uniform within the hard-wall trap of radius $R$, apart from a Gaussian-like hole centered at $\bm{r}_v$ and having width $\sigma_a$. As regards the phase field 
\begin{equation}
    \label{eq:theta_a}
    \theta_a= \arctan\left(\frac{y-y_v}{x-x_v}\right)-\arctan\left(\frac{y-y_v^\prime}{x-x_v^\prime}\right),
\end{equation}
it features singularities at $|\bm{r}_v|<R$ and at $|\bm{r}_v^\prime|>R$,
where $(x_v,\,y_v):=\bm{r}_v^\prime=\bm{r}_v R^2/r_v^2$. The first one coincides with the center of the Gaussian-like hole and thus represents an actual vortex, while the second one is the so called ``image vortex'', originating from the presence of the circular boundary, and ensuring that the $a$-component velocity field $\bm{v}_a\propto \nabla\theta_a$ is purely tangential to it. 

In the same spirit of Ref. \cite{Richaud2021}, we assume that $b$-component bosons are described by a Gaussian wavefunction
\begin{equation}
\label{eq:psi_b}
\psi_b(\bm r) =  \left(\frac{N_b}{\pi\sigma_b^2}\right)^{1/2}e^{-|\bm r - \bm r_c(t)|^2/2\sigma_b^2}e^{i\bm r\cdot \bm \alpha(t)}
\end{equation}
but, as opposed to our previous study, in this work we take into account the possible \textit{relative motion} between the vortex and its massive core. Therefore, in general, $\bm{r}_v\neq\bm{r}_c$, a circumstance resulting in an increased number of time-dependent variational parameters and thus in a more accurate modelling of the physical problem. 

Substituting the time-dependent variational ansatzes (\ref{eq:psi_a}) and (\ref{eq:psi_b}) in the field Lagrangian
$$
  L_{\rm GP}[\psi_a,\,\psi_b]=\sum_{i=a,\,b} \left[ \frac{i\hbar}{2}\int \mathrm{d}^2 r\left(\psi_i^*\frac{\partial\psi_i}{\partial t} - \frac{\partial \psi_i^*}{\partial t }\psi_i\right) -\right.
$$ 
$$
 \left.  \int \mathrm{d}^2r\left(\frac{\hbar^2}{2m_i}|\bm \nabla \psi_i|^2 +V_{\rm tr}|\psi_i|^2 + \frac{g_i}{2}|\psi_i|^4\right) \right ] -
$$
\begin{equation}
\label{eq:L_GP}
g_{ab} \int\mathrm{d}^2r\, |\psi_a|^2|\psi_b|^2
\end{equation}	
[generating Euler-Lagrange equations (\ref{eq:H_GPE})
], and after integrating out the fields' degrees of freedom, one obtains the following effective Lagrangian, up to additive constant terms.
$$
  L=\hbar n_a \pi \dot{\bm r}_v\times {\bm r}_v \cdot \hat{z}- \frac{n_a \pi\hbar^2}{m_a}\log\left[1-\left(\frac{r_v}{R}\right)^2\right] +
 $$
\begin{equation}
    \label{eq:L_tot}
   \frac{1}{2}N_bm_b \dot{\bm{r}}_c^2 - g_{ab} n_a N_b \sigma_{ab} |\bm{r}_v-\bm{r}_c|^2   
\end{equation}
where the density-density repulsion energy [see the term $\propto g_{ab}$ in Eq. (\ref{eq:L_GP}) and appendix \ref{EabComputation}] has been expanded to the second order in the displacement $|\bm{r}_v-\bm{r}_c|  $ and where $\sigma_{ab}=\sigma_a^2/(\sigma_a^2+\sigma_b^2)^2$. We remark that, as opposed to the massive-point vortex model presented in Refs. \cite{Richaud2020,Richaud2021,Richaud2022r_k,Richaud2022Collisions}, in this model, vortices and massive cores are described by different sets of dynamical variables.

The associated Euler-Lagrange equations
\begin{equation}
    \label{eq:EL_vortices}
    2\pi h\,  \dot{\bm{r}}_v \times \hat{z} =\frac{h^2}{m_a}\frac{\bm{r}_v}{R^2-r_v^2} +4\pi g_{ab}N_b \sigma_{ab} (\bm{r}_c-\bm{r}_v),
\end{equation}
\begin{equation}
\label{eq:EL_cores}
m_bN_b \ddot{\bm{r}}_c =- 2g_{ab} \sigma_{ab} n_a N_b (\bm{r}_c-\bm{r}_v)
\end{equation}
show that the vortex velocity $\dot{\bm r}_v$ not only depends on the vortex position $\bm{r}_v$ (as it is customarily the case for a standard superfluid vortex \cite{Kim2004}), but also on the position $\bm{r}_c$ of its massive core, which is in turn subjected to a recoil force towards the vortex center. On the other hand, the motion equation for $\bm{r}_v(t)$ is of first order, in analogy with well-established (massless) point-vortex models \cite{Lin1943,Onsager1949,Hess1967,Kim2004,Middelkamp2010,Middelkamp2011,Torres2011,Torres2011vortex, Navarro2013,
Simula2014,Murray2016,Kim2016,Groszek2018}, but unlike the massive-point vortex model discussed in Refs. \cite{Richaud2020,Griffin2020,Richaud2021,Richaud2022r_k,Richaud2022Collisions,Caldara2023} 
where the time evolution of vortex positions is governed by second-order motion equations.
Eventually, it is worth remarking that, upon multiplying both sides of Eq. (\ref{eq:EL_vortices}) for the 2D number density $n_a$, all terms have the dimensions of a force. In this optics, one can conclude that the (massless) vortex moves to ensure that the total force acting on it is zero, which is exactly the Magnus effect \cite{Richaud2022r_k}.
For $g_{ab}\to \infty$ and $\bm{r}_c\to\bm{r}_v$, the two equations reduce to the previous model ones, as in Ref. \cite{Richaud2021}.

\subsection{Hamiltonian description}

We introduce
the equivalent description of the system dynamics within the Hamiltonian framework, which will prove particularly convenient for the study of the system's normal modes
and of their properties (see Sec. \ref{sec:Normal_modes}). 
Preliminarily, we compute the canonical momenta of the vortex 
\begin{equation}
    \label{eq:p_v}
    \bm{p}_v =\partial L/\partial \dot{\bm{r}}_v= \frac{n_a h}{2}\bm{r}_v\times \hat{z}
\end{equation}
and of the massive core
\begin{equation}
    \label{eq:p_c}
    \bm{p}_c =\partial L/\partial \dot{\bm{r}}_c= m_b N_b \dot{\bm{r}}_c,
\end{equation}
and observe that, while the second momentum exhibits the expected linear dependence on the velocity, the first one is structurally different as the momentum components of the vortex depend on the vortex-position coordinates. The presence of constraints between canonical variables 
apparently prevents the transition to the Hamilton formalism. This pathology is
removed by applying the Dirac procedure for constrained Hamiltonian systems
\cite{Hanson1976,Dirac1950} which allows one to derive new Dirac-Poisson brackets incorporating the dynamical  constraints. In the current case, the new brackets read
$$
 \{A,\,B\}= \frac{1}{h n_a}\left(\frac{\partial A}{\partial x_v}\frac{\partial B}{\partial y_v} - \frac{\partial B}{\partial x_v}\frac{\partial A}{\partial y_v}\right) +
$$
\begin{equation}
  \frac{\partial A}{\partial x_c}\frac{\partial B}{\partial p_{x_c}}  - \frac{\partial B}{\partial x_c}\frac{\partial A}{\partial p_{x_c}}  +  \frac{\partial A}{\partial y_c}\frac{\partial B}{\partial p_{y_c}}  - \frac{\partial B}{\partial y_c}\frac{\partial A}{\partial p_{y_c}},  
\label{DB}
\end{equation}
with the Hamiltonian
$$
  H=\frac{h^2 n_a}{4 \pi m_a} \log\left[1-\left(\frac{r_v}{R}\right)^2\right] +
$$
\begin{equation}
    \label{eq:H}
    \frac{\bm{p}_c^2}{2 m_b N_b} + g_{ab} \sigma_{ab} n_a N_b |\bm{r}_v -\bm{r}_c|^2.
\end{equation}
The effect of Dirac's method is clearly visible in formula (\ref{DB}) where $x_v$ and $y_v$ have the status of canonically conjugate variables. This property, extended to many-vortex systems, allows for the derivation of Helmholtz-Kirchhoff
equations within the Hamilton picture \cite{Aref1983}.

Upon defining the vector of dynamical variables 
$\bm{z}=(x_v,\,y_v,\,x_c,\,y_c,\,p_{x_c},p_{y_c})$, the Hamilton motion equations are given by 
\begin{equation}
    \label{eq:Hamilton_equations}
    \dot{\bm{z}}=\{\bm{z},\,H\}
\end{equation}
which are indeed equivalent to Eqs. (\ref{eq:EL_vortices})-(\ref{eq:EL_cores}). 
The canonical angular momenta for the vortex and the mass read
\begin{equation}
    \label{eq:l_v}
    \bm{l}_v=\bm{r}_v\times \bm{p}_v = -\frac{h n_a}{2} r_v^2 \,\hat{z}
\end{equation}
\begin{equation}
    \label{eq:l_c}
    \bm{l}_c=\bm{r}_c\times \bm{p}_c = m_b N_b\, \bm{r}_c\times \dot{\bm{r}}_c,
\end{equation}
respectively.
Due to the planar character of the mass and vortex motions, the two angular momenta
$\bm{l}_v$ and $\bm{l}_c$ are aligned with the $z$ axis as well as the total angular momentum $\bm{L}=L_z \hat{z} = \bm{l}_v+\bm{l}_c$. Then, one easily verifies that $\{ L_z,\,H\}= 0$, a conservation property which comes from the rotational symmetry of the system. 

We remark that, while $\bm{l}_c$
corresponds to the expectation value 
$\langle \hat{\bm{L}} \rangle_{\psi_b}$ 
of the angular-momentum 
$\hat{\bm{L}}$ evaluated with respect to ansatz (\ref{eq:psi_b}), one can easily verify that $\bm{l}_v$ differs from $\langle \hat{\bm{L}}\rangle_{\psi_a} = h n_a (R^2-r_v^2)/2\, \hat{z}$ [computed using ansatz (\ref{eq:psi_a}) in the limit $\sigma_a\to 0$] due to a constant additive term $h n_a R^2/2$. 
This difference does not constitute an issue
since Lagrangians are known to be defined up total time derivatives. By defining the Lagrangian $L^\prime:=L+\hbar n_a\pi R^2\, (\mathrm{d}\theta_v/\mathrm{d}t)$ one obtains the same motion equations of Lagrangian (\ref{eq:L_tot}) together with the expected vortex angular momentum $\bm{l}_v =\langle \hat{\bm{L}}\rangle_{\psi_a}$. 

\subsection{Uniform circular orbits}
\label{sub:Uniform_circular_orbits}
\begin{figure}[h!]
    \centering
    \includegraphics[width=0.5\columnwidth]{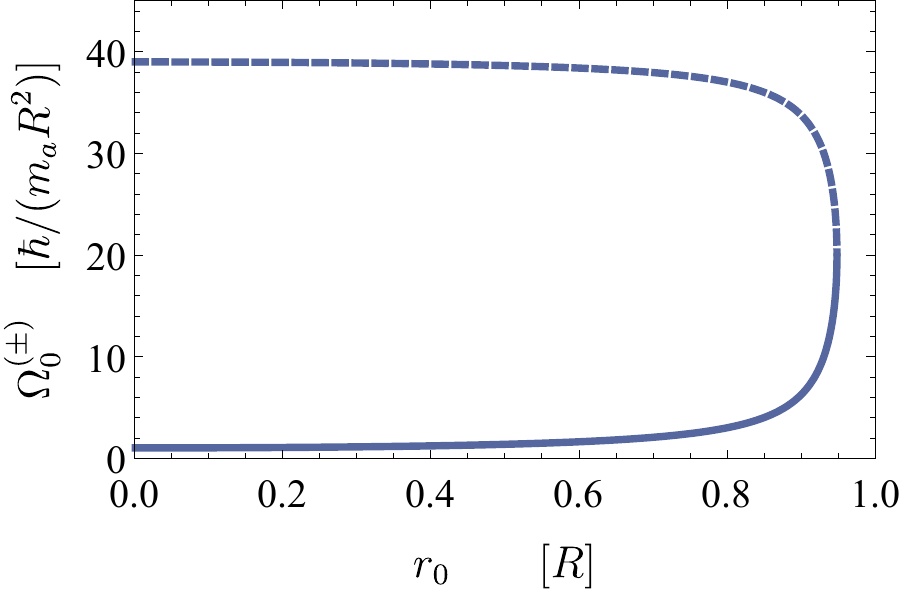}
    \caption{Functional dependence of Eqs. (\ref{eq:Omega_plus}) (dashed line) and (\ref{eq:Omega_minus}) (solid line) on the normalized radial distance $r_0$ for $\mu=0.05$. }
    \label{fig:Omega_plus_minus}
\end{figure}

Equations (\ref{eq:EL_vortices})-(\ref{eq:EL_cores}) admit a notable class of solutions, such that both the vortex and its massive core exhibits a uniform circular motion with angular frequency $\Omega$. In this case, the trajectory can be parametrized as 
\begin{equation}
x_v = \tilde{r}_v \cos (\Omega t), \,
y_v = \tilde{r}_v \sin (\Omega t),\,
\label{C1}
\end{equation}
\begin{equation}
x_c = \tilde{r}_c \cos (\Omega t), \,
y_c = \tilde{r}_c \sin (\Omega t), \,
\label{C2}
\end{equation}
where $\tilde{r}_v$ and $\tilde{r}_c$  are time-independent quantities depending on $\Omega$ (and on the other model parameters) whose defining equations are discussed in Appendix \ref{app1}. Although the analytic expression of the orbits' radii $\tilde{r}_v(\Omega)$ and $\tilde{r}_c(\Omega)$ is rather complex, it is possible to point out some remarkable properties. Their ratio, 
\begin{equation}
    \label{eq:r_c_over_r_v}
    \frac{\tilde{r}_c}{\tilde{r}_v}=\frac{2 g_{ab} n_a \sigma_{ab} }{2 g_{ab} n_a \sigma_{ab} - m_b\Omega^2} 
\end{equation}
is always greater than $1$ because the massive core is subject to an effective centrifugal force and tends to $1^+$ for $g_{ab}\to +\infty$ (highly-immiscible species). In this limit, as shown in Appendix \ref{app1}, one recovers the results of Ref. \cite{Richaud2021}, i.e.
\begin{equation}
\label{eq:Omega_plus}
\Omega_0^{(+)} =
\frac{1+\sqrt{1-2\mu/(1-r_0^2)}}{\mu}  
\end{equation}
and 
\begin{equation}
\label{eq:Omega_minus}
\Omega_0^{(-)} =  \frac{2/(1-r_0^2)}{1+\sqrt{1-2\mu/(1-r_0^2)}}
\end{equation}
here expressed in units of $\hbar/(m_aR^2)$ (frequency) and $R$ (length), where $\mu=m_aN_a/(m_bN_b)$ represents the mass imbalance, $r_0 := {\tilde r}_0 /R$ is the effective radius, and $\tilde{r}_v , \tilde{r}_c = {\tilde r}_0$ (see also Fig. \ref{fig:Omega_plus_minus}). Moreover, in the massless limit $N_b\to 0$, $\Omega_0^{(+)}$ diverges (and thus turns unphysical), while  $\Omega_0^{(-)}$ reduces to $\Omega=\hbar/[m_a(R^2-r_v^2)]$, the well-known frequency for a vortex inside a circular boundary \cite{Kim2004}.

\section{Normal modes}
\label{sec:Normal_modes}
\begin{figure}[h!]
    \centering
    \includegraphics[width=0.5\columnwidth]{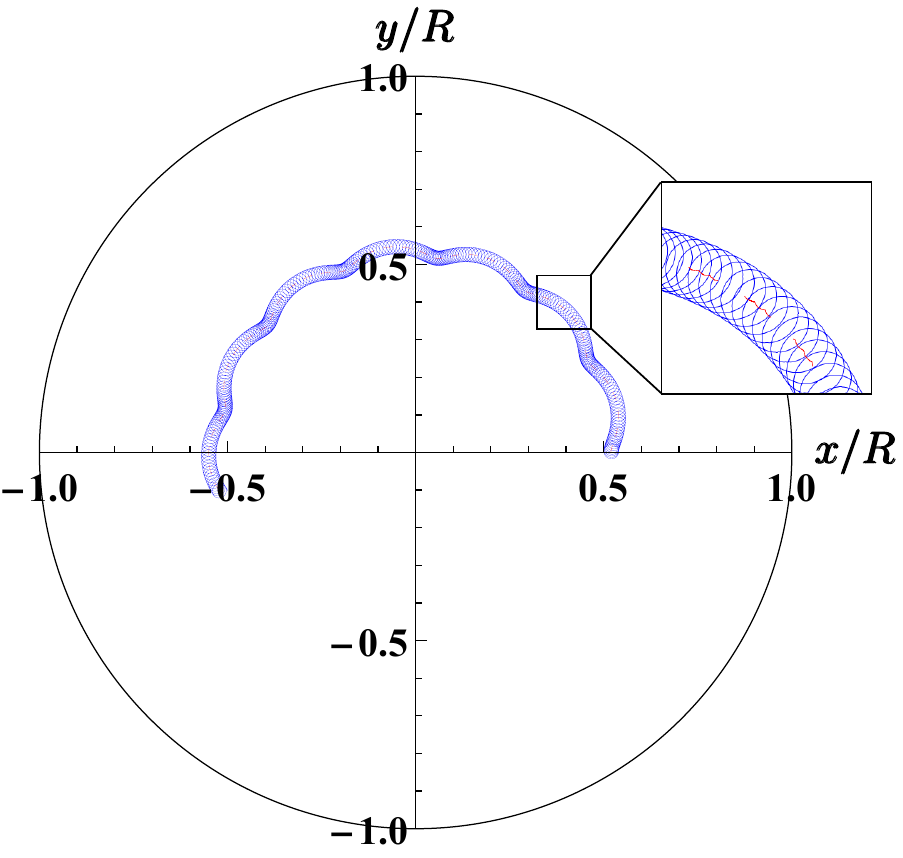}
    \caption{Numerical solution of Eqs. (\ref{eq:EL_vortices})-(\ref{eq:EL_cores}). Solid blue (red dashed) line corresponds to the trajectory of the vortex (massive core). The inset allows one to appreciate the rapid precession of the vortex around its massive core. The frequency of their rapid precession can be estimated by Eq. (\ref{eq:Asymptotic_omega_1}). The microscopic parameters are those of a $^{23}\mathrm{Na}+^{39}\mathrm{K}$-mixture \cite{Richaud2019}. In particular, the interspecies scattering length is taken to be $a_{12}=24\,a_0$ (where $a_0$ is the Bohr radius), $d_z=0.4\,\mu m$, $\sigma_a= 5\,\mu m$, $\sigma_b= 4\,\mu m$, $N_a=10^5$, $N_b=5\times 10^3$, $R=50\,\mu m$. The dynamics is triggered by perturbing the overall uniform circular orbit corresponding to a radius of $R/2$. The dynamics has been simulated for $\Delta t=2\,s$.}
    \label{fig:Trajectory}
\end{figure}

Small perturbations of the uniform circular orbits discussed in Sec. \ref{sub:Uniform_circular_orbits} trigger small-amplitude oscillations and result in more complex trajectories. A typical example is presented in Fig. \ref{fig:Trajectory}, where one can appreciate, besides an overall motion of precession around the trap center at frequency $\sim \Omega$, two additional modes: a radial oscillation of the vortex-massive-core complex, and a circular motion of precession of the vortex around its massive core. While the former has been already described in Ref. \cite{Richaud2021} and explained to be the hallmark of the core's inertial mass, the latter consists in a relative motion between the quantum vortex and its massive core and could not be predicted within previous approaches \cite{Richaud2020,Richaud2021}, which relied on the approximation $\bm{r}_v \equiv \bm{r}_c$. Interestingly, the direction of the precession of the vortex around its massive core (clockwise in Fig. \ref{fig:Trajectory}) is \textit{opposite} with respect to that of the vortex-massive-core complex around the tarp center (anticlockwise in Fig. \ref{fig:Trajectory}). 

\subsection{Linear response}
\label{LINR}

The dynamical stability of the uniform circular orbits discussed in Sec. \ref{sub:Uniform_circular_orbits}, as well as the analysis of the system's normal modes is carried out according to the standard techniques from Dynamical Systems Theory. One starts from Lagrangian  (\ref{eq:L_tot}) written in the laboratory reference frame $xOy$, and then switches to a reference frame $XOY$, rotating at angular frequency $\Omega$, by means of the transformations 
\begin{equation}
x_i= X_i C(t)-Y_i S(t), \,\,
y_i =X_i S(t) + Y_i C(t),
\label{rot}
\end{equation} 
where $C(t)=\cos(\Omega t)$, $ S(t)=\sin(\Omega t)$ and $ i=(v,\,c)$ (the velocities transform accordingly). The Hamiltonian $H_{\rm rot}$ of the system in this new reference frame is readily  computed (notice that it is formally equivalent to $H-\Omega L_z$). The six Hamilton equations are written as $\dot{\bm Z}=\mathbb{E} (\nabla H_{\rm rot})^T$
[their explicit expression is given in Appendix (\ref{app2})], where $\bm Z=(X_v,\,Y_v,\,X_c,\,Y_c,\,P_{X},\,P_{Y})^T$ is the vector of dynamical variables, i.e. the generalized coordinates and their momenta in the rotating frame, and 
\begin{equation}
    \mathbb{E}=\left(
\begin{array}{cccccc}
 0 & \frac{1}{h n_a} & 0 & 0 & 0 & 0 \\
 -\frac{1}{h n_a} & 0 & 0 & 0 & 0 & 0 \\
 0 & 0 & 0 & 0 & 1 & 0 \\
 0 & 0 & 0 & 0 & 0 & 1 \\
 0 & 0 & -1 & 0 & 0 & 0 \\
 0 & 0 & 0 & -1 & 0 & 0 \\
\end{array}
\right)
\end{equation}
is a suitable symplectic matrix (notice that $x_v$ and $y_v$ are canonically conjugate, as it is customarily in classical nonviscous hydrodynamics \cite{Onsager1949}, while the massive core is associated to actual pairs of coordinates and conjugate momenta). 
The fixed points of this dynamical system correspond to the uniform circular orbits discussed in Sec. \ref{sub:Uniform_circular_orbits}. Then one computes the Hessian matrix $\mathbb{H}$ and the Jacobian matrix $\mathbb{J}=\mathbb{E}\mathbb{H}$ associated to the dynamical system 
(see Appendix \ref{app2})
and evaluates the latter at the previously found fixed points (see Sec. IV of Ref. \cite{Richaud2018} for a detailed explanation). The eigenvalues of $\mathbb{J}$ characterize the stability of the fixed point (namely the uniform circular orbit in the laboratory frame): if they are purely imaginary (have a non-zero real part), then the fixed point, is dynamically stable (unstable). The imaginary parts $\omega_j$ of the eigenvalues $\lambda_j=\pm i \omega_j$ represent the eigenfrequencies of the system's normal modes (we remark that a pair of eigenvalues is always zero, a circumstance which originates from presence of a conserved quantity, namely $\bm{l}\cdot\hat{z}$). 

In the following, we use the symbol $\omega_1$ to denote the frequency of the very rapid precession of the vortex around its massive core (see the inset of Fig. \ref{fig:Trajectory}) and the symbol $\omega_2$ to indicate the frequency of the radial oscillations of the vortex-massive-core complex (see the main panel of Fig. \ref{fig:Trajectory}). 

\subsection{Eigenfrequencies of a vortex-massive-core at the trap center}
While for generic values of $R$, $g_{ab}$, and $N_b$, the eigenfrequencies $\omega_1$ and $\omega_2$ do not
exhibit simple analytic expressions, it is instructive to discuss their properties for $\tilde{r}_c=\tilde{r}_v=0$. In this case, in fact, the vortex and its massive core both lie at the trap center, a circumstance which implies that the image vortex is infinitely far and hence does not affect the system dynamics. Under these hypotheses, the eigenfrequencies can be written as:

\begin{equation}
    \label{eq:omega_1_0}
        \omega_1^0= \frac{\gamma_{ab} \sqrt 2}{h m_b}
    \sqrt{ 1 + \frac{h^2 n_a}{\gamma_{ab} N_b} 
    +\sqrt{1+ 2 \frac{h^2 n_a}{\gamma_{ab} N_b} 
    }},
\end{equation}

\begin{equation}
    \label{eq:omega_2_0}
    \omega_2^0= \frac{\gamma_{ab} \sqrt 2}{h m_b}
    \sqrt{ 1 + \frac{h^2 n_a}{\gamma_{ab} N_b} 
    -\sqrt{1+ 2 \frac{h^2 n_a}{\gamma_{ab} N_b} 
    }},
\end{equation}
with $\gamma_{ab}= g_{ab} \sigma_{ab} m_b N_b$.
Interestingly, in the limit of high immiscibility $g_{ab}\to +\infty$, the rapid-precession frequency $\omega_1^0$ diverges and thus turns unphysical, while the oscillation frequency of the overall transverse oscillations $\omega_2$ tends to $h n_a/(m_b N_b)$. The latter value exactly corresponds to the small-oscillation frequency computed in Ref. \cite{Richaud2021}.

In the limit of large $N_b$,
describing a very heavy massive core, provides another interesting case. In this circumstance, the large inertia of the massive core dominates the dynamics of the overall system, and transverse radial oscillations of the vortex-massive-core complex are prohibited, i.e. $\omega_2=0$. On the other hand, one can verify that,
for $N_b$ large enough
\begin{equation}
    \label{eq:Asymptotic_omega_1}
 \omega_1^* \simeq
 \frac{2 g_{ab} \sigma_{ab} N_b}{h}
\end{equation}
which is indeed a finite quantity. Notice, in this regard, that the latter value may have been derived directly from the inspection of Eq. (\ref{eq:EL_vortices}) and, more specifically, by applying the classical counterpart of the Born-Oppenheimer approximation to it. Observing Fig. \ref{fig:Trajectory}, in fact, one can notice the presence of a very rapid precession of the vortex around its massive core. Since, during a full precession cycle, the massive core can be regarded as stationary, one can neglect the time dependence of $\bm{r}_c$. Under this approximation, (and in the limit of large $R$), the effective dynamical equation of the vortex is thus
\begin{equation}
\label{eq:Simplified_Lag_eq_r_v}
      2\pi h\,  \dot{\bm{r}}_v \times \hat{z} = 4\pi g_{ab}N_b \sigma_{ab} (\bm{r}_c-\bm{r}_v),
\end{equation}
which depends on the only unknown $\bm{r}_v(t)$. 
The solution of this equation, formally equivalent to that of a charge in a constant magnetic field, corresponds to a uniform circular motion centered at $\bm{r}_c$ whose precession frequency exactly corresponds to the limit (\ref{eq:Asymptotic_omega_1}) derived above within the linear-response framework. We remark that, in this regime, the eigenfrequency $\omega_1$ \emph{linearly} depends on the coupling constant $g_{ab}$.

With reference to Fig. \ref{fig:Trajectory}, for example, over a simulated time interval of $2\, s$, the vortex performs 279 revolutions around its massive core, corresponding to a rate of 877 $rad/s$. This  ``experimental" data deviates less than $5\%$ from the estimate one can obtain from Eq. (\ref{eq:Asymptotic_omega_1}), namely 836 $rad/s$, and less than $2\%$ from the estimate provided by Eq. (\ref{eq:omega_1_0}), i.e. 861 $rad/s$. 

\subsection{Eigenfrequencies of an off-centered vortex-massive-core complex}
\label{sub:Eigenfrequencies_off_centered_vortex}
As already mentioned, the eigenfrequencies $\omega_1$ and $\omega_2$ can be written in closed form [see Eqs. (\ref{eq:omega_1_0}) and (\ref{eq:omega_2_0})] only upon neglecting the presence of the circular trap. Already within this approximation, and, surprisingly, also upon introducing the further simplification that the massive core is stationary with respect to the fast precession of the massless vortex [see Eq.  (\ref{eq:Asymptotic_omega_1})], the analytical estimates of $\omega_1$ are in good quantitative agreement with the results extracted from the full solution of the time-dependent dynamical equations. 

\begin{figure}[h!]
    \centering
    \includegraphics[width=1\columnwidth]{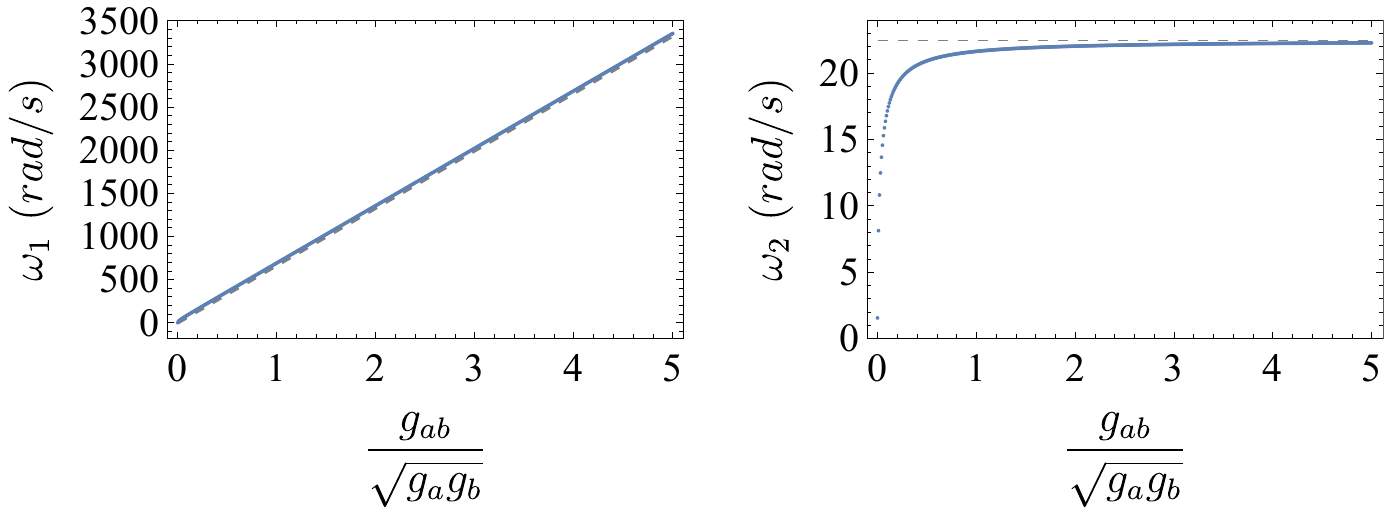}
    \caption{Functional dependence of the two eigenfrequencies $\omega_1$ (left panel) and $\omega_2$ (right panel) with respect to the interspecies coupling $g_{ab}$. Blue dots correspond to numerical results, while gray dashed line corresponds to analytical predictions 
    Eq. (\ref{eq:Asymptotic_omega_1}) 
    [Eq. (\ref{eq:omega_2_*})]
    for the left (right) panel. Microscopic model parameters 
    are those used in Fig. \ref{fig:Trajectory}.   }
    \label{fig:Eigenfrequencies_vs_gab}
\end{figure}

In the following, we present a full-fledged analysis of the system eigenmodes, computed numerically taking into account the finite size ($R$) of the superfluid sample as well as the presence of the hard-wall circular boundary. Figure \ref{fig:Eigenfrequencies_vs_gab} illustrates the functional dependence of $\omega_1$ and $\omega_2$ with respect to the intercomponent repulsive interaction $g_{ab}$. As visible from the left panel, $\omega_1$ features a persistently linear dependence on $g_{ab}$, signaling the fact that neither the overall motion of precession of the 
quantum-vortex-massive-core complex nor its transverse oscillations significantly affect this eigenmode, which is, instead, governed mainly by the effective attraction between the vortex and its massive core [see Eq. (\ref{eq:Simplified_Lag_eq_r_v})].

As regards the right panel of Fig. \ref{fig:Eigenfrequencies_vs_gab}, one can notice a strong dependence of $\omega_2$ on the interspecies coupling. In the limit $g_{ab}\to 0^+$, in fact, $\omega_2$ tends to zero, a circumstance signaling the fact that the quantum vortex, upon being disconnected from its massive core, looses its inertia and ceases to oscillate. In the opposite limit, i.e. for $g_{ab}\to +\infty$, $\omega_2$ asymptotically approaches the limiting value
\begin{equation}
\label{eq:omega_2_*}
    \omega_2^*=\frac{\hbar}{m_aR^2} \frac{2}{\mu} \sqrt{1-\mu\frac{2-\tilde{r}_0^2}{(1-\tilde{r}_0^2)^2}}
\end{equation}
where $\mu=m_bN_b/(m_a n_a \pi R^2)$ is the mass ratio, and $\tilde{r}_0$ is the position (given in units of $R$) of a massive core tightly bound to the center of its hosting quantum vortex. This expression was indeed derived in Ref. \cite{Richaud2021} upon assuming $\bm{r}_v\equiv \bm{r}_c$, a constraint which is, in turn, consistent with the condition $g_{ab}\to +\infty$. 
In this case, in fact, the system tends to minimize the overlap between $|\psi_a|^2$ and $|\psi_b|^2$, thus resulting in a tight confinement of the massive core within its hosting quantum vortex. We remark that, unexpectedly, the oscillation frequency of the massive vortex depends on the spring constant $\propto g_{ab}$ because the presence of the image vortex can be interpreted as an external force.

Let us now turn to the analysis of the dependence of $\omega_1$ and $\omega_2$ on the number of $b$-component particles.

\begin{figure}[h!]
    \centering
    \includegraphics[width=1\columnwidth]{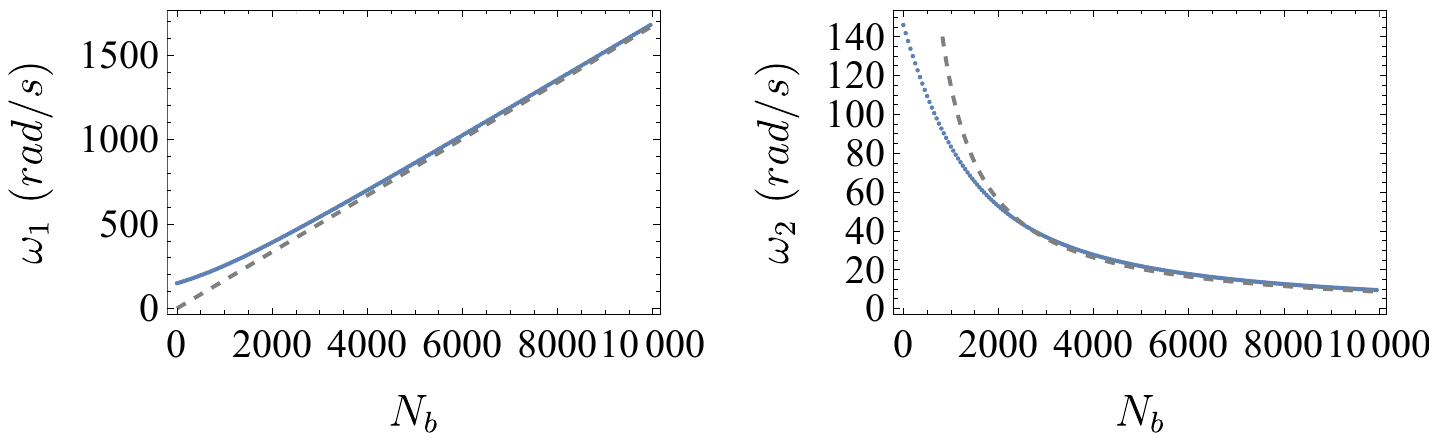}
    \caption{Functional dependence of the two eigenfrequencies $\omega_1$ (left panel) and $\omega_2$ (right panel) with respect to the number of core particles $N_b$. Blue dots correspond to numerical results, while gray dashed lines corresponds to analytical predictions [Eq. (\ref{eq:Asymptotic_omega_1}) for the left panel, and Eq. (\ref{eq:omega_2_*}) for the right panel]. Microscopic model parameters correspond to the ones used in Fig. \ref{fig:Trajectory} and Fig. \ref{fig:Eigenfrequencies_vs_gab}.   }
    \label{fig:Eigenfrequencies_vs_Nb}
\end{figure}

The left panel of Fig. \ref{fig:Eigenfrequencies_vs_Nb} shows that $\omega_1$ linearly depends on $N_b$ for large values of $N_b$, the oblique asymptote being given by Eq. (\ref{eq:Asymptotic_omega_1}). Interestingly, the numerically-obtained data deviates from the aforementioned asymptote for small values of $N_b$. One can verify that
\begin{equation}
\label{eq:omega_harmonic_oscillator}
    \lim_{N_b\to 0} \omega_1^0 = \sqrt{\frac{2g_{ab} \sigma_{ab} n_a}{m_b}},
\end{equation}
a quantity which basically corresponds to the harmonic-oscillator frequency associated to Eq. (\ref{eq:EL_cores}) under the approximation that $\bm{r}_v$ is constant. Further comments will follow in Sec. \ref{sub:Light_core}. As regards $\omega_2$ (right panel of Fig. \ref{fig:Eigenfrequencies_vs_Nb}), one can see that its functional dependence on $N_b$ is well captured by Eq. (\ref{eq:omega_2_*}) in the large-$N_b$ limit (blue dots are essentially on top of the gray dashed line). Nevertheless, as opposed to the behaviour predicted by Eq. (\ref{eq:omega_2_*}), $\omega_2$ does not diverge for $N_b\to 0$, but approaches a limiting value very close to quantity (\ref{eq:omega_harmonic_oscillator}). This unexpected circumstance will be discussed in more detail in Sec. \ref{sub:Light_core}.

\subsection{Light-core vortices and hybridization of the eigenmodes}
\label{sub:Light_core}
\begin{figure}[h!]
    \centering
    \includegraphics[width=1\columnwidth]{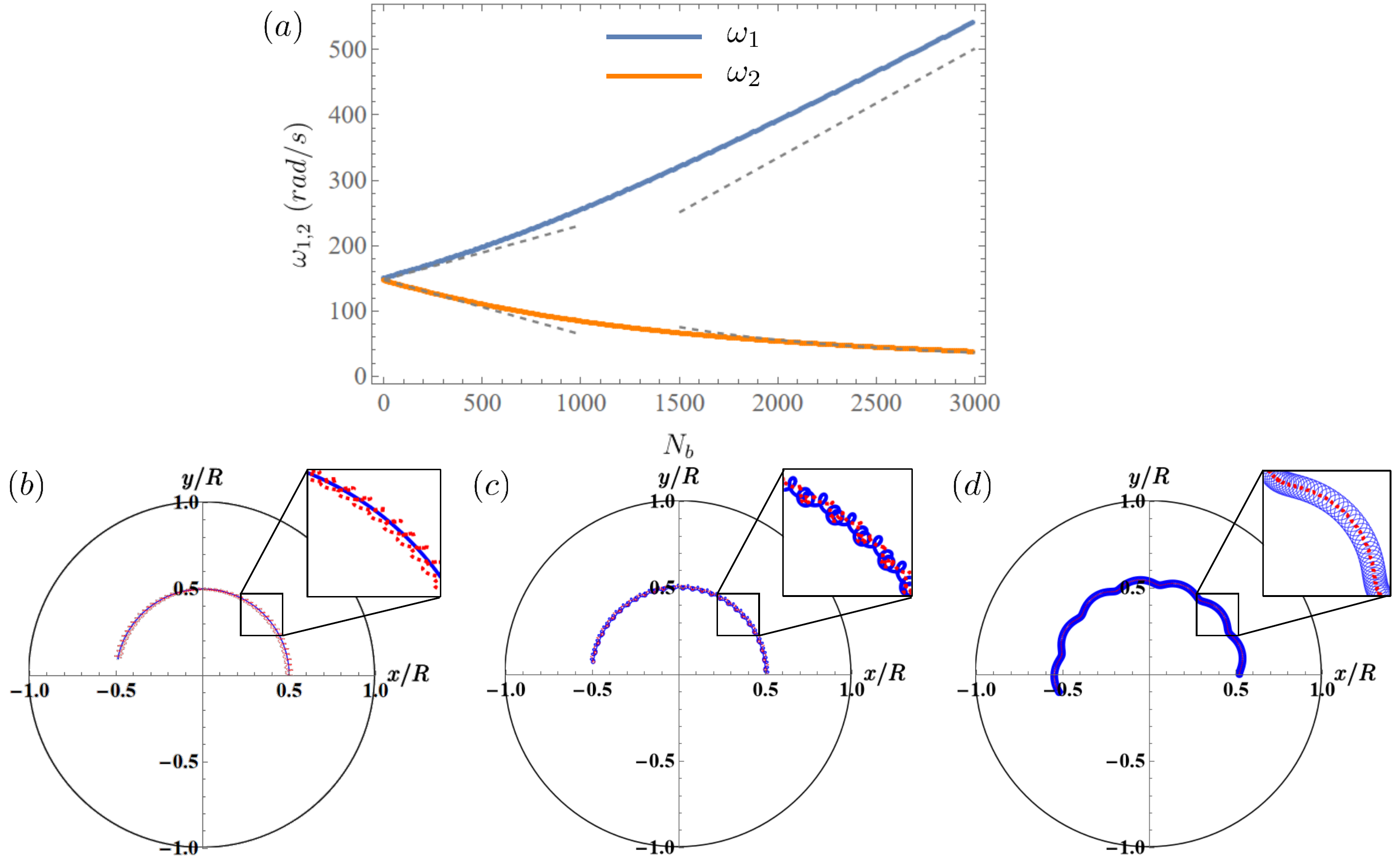}
    \caption{Panel (a): Functional dependence of the two eigenfrequencies $\omega_1$ (blue line) and $\omega_2$ (orange line) on the number of core particles $N_b$. They were computed numerically taking into account the finite size ($R$) of the superfluid sample. Gray dashed lines correspond to some analytical estimates: the upper left gray line corresponds to Eq. (\ref{eq:omega_1_left}), lower left gray line corresponds to Eq. (\ref{eq:omega_2_left}), upper right gray line corresponds to Eq. (\ref{eq:Asymptotic_omega_1}), while the lower right gray line corresponds to Eq. (\ref{eq:omega_2_*}). Panels (b), (c), and (d) illustrate the detailed structure of the normal modes for three different values of the mass ratio $\mu$. Panel (b) corresponds to $\mu=10^{-5}$, a regime where the very light massive core basically oscillates within the hosting vortex, which plays the role of an isotropic 2D harmonic trap (and whose trajectory is almost unaffected by the presence of the massive core). Panel (d) corresponds to $\mu=1.7\times 10^{-2}$, a regime where the quantum vortex exhibits a very rapid precession around its massive core and the whole complex features radial oscillations. Panel (c) corresponds to $\mu=8.5\times 10^{-2}$, a regime where the aforementioned normal modes get hybridized and result in rather complex trajectories, which cannot be easily described. The same microscopic model parameters of Fig. \ref{fig:Trajectory} were used, with $N_b=1,\,10^3,\,5\times 10^3$ for panel (b), (c), and (d), respectively. }
    \label{fig:Hybrid_normal_modes}
\end{figure}

As already mentioned, in the limit of very light massive cores ($N_b\to 0$), both $\omega_1$ and $\omega_2$ basically tend to quantity (\ref{eq:omega_harmonic_oscillator}). 
In this circumstance, in fact, both eigenfrequencies reduce to those ones of an isotropic 2D harmonic oscillator, constituted by the light massive core trapped within a stationary harmonic potential.
It is worth remarking that, if the center of the harmonic potential is not stationary, but orbits at a rate $\Omega$, then the original system's eigenfrequencies $\omega_1=\omega_2$ given by Eq. (\ref{eq:omega_harmonic_oscillator}) modify as follows: 
$$
\omega_1\to\omega_1^\prime :=\omega_1+ \Omega, \quad \omega_2\to\omega_2^\prime=\omega_2-\Omega. 
$$
Upon increasing the value of $N_b$, the (quasi-)degeneracy of $\omega_1$ and $\omega_2$ is lifted (see Fig. \ref{fig:Hybrid_normal_modes}) and, up to the first order in $N_b$, the two eigenfrequencies 
\begin{equation}
    \label{eq:omega_1_left}
    \omega_1 \approx \sqrt{\frac{2 g_{ab} \sigma_{ab} n_a}{m_b}} +\Omega+\frac{g_{ab} \sigma_{ab} }{h}N_b +\mathcal{O}(N_b^2)
\end{equation}
\begin{equation}
    \label{eq:omega_2_left}
    \omega_2 \approx \sqrt{\frac{2 g_{ab} \sigma_{ab} n_a}{m_b}}-\Omega -\frac{g_{ab} \sigma_{ab} }{h}N_b +\mathcal{O}(N_b^2)
\end{equation}
symmetrically depart from quantity (\ref{eq:omega_harmonic_oscillator}) [see also the two dashed lines on the left-hand side of Fig. \ref{fig:Hybrid_normal_modes} (a)]. We remark that, in this limit, the normal modes consist only in the oscillations of the light massive core around a center orbiting at frequency $\Omega$, while the quantum vortex is basically unaffected [see Fig. \ref{fig:Hybrid_normal_modes} (b)]. 


On the other hand, for large values of $N_b$, $\omega_1$ and $\omega_2$ are well approximated by Eq. (\ref{eq:Asymptotic_omega_1}) and Eq. (\ref{eq:omega_2_*}) respectively [see gray dashed lines on the right-hand side of Fig. \ref{fig:Hybrid_normal_modes} (a)]. In this limit, the structure of the normal modes are very different: the eigenmode associated to $\omega_1$ consists in a very rapid precession of the quantum vortex around its massive core [see Eq. (\ref{eq:Simplified_Lag_eq_r_v}) and the relevant discussion], while that associated to $\omega_2$ consists in radial oscillations of the quantum-vortex-massive-core complex [see Fig. \ref{fig:Hybrid_normal_modes} (c)].

As visible from Fig. \ref{fig:Hybrid_normal_modes} (a), there exists a range of values of $N_b$ (in this case $N_b\sim 1500$) where the analytical estimates of $\omega_1$ and $\omega_2$ valid for small (gray lines on the left-hand side) and large (gray lines on the right-hand side) values of $N_b$ get (pairwise) similar. This implies that the two pairs of previously discussed normal modes ``hybridize", meaning that their shape [see Fig. \ref{fig:Hybrid_normal_modes} (b)] inherits some properties of both parent eigenmodes. More specifically, the resulting eigenmodes involve the oscillations of both the quantum vortex and its massive core, thus resulting in rather complex trajectories, which, to the best of our knowledge, elude further analytical description.

\section{Gross-Pitaevskii results}
\label{sec:GP_results}

We compare the predictions of the analytical model discussed in Sec. \ref{sec:Model} and Sec. \ref{sec:Normal_modes} with numerical simulations of coupled Gross-Pitaevskii equations (\ref{eq:H_GPE}), which, in turn, well capture the essential physics of a Bose-Bose mixture at very low temperature. 
\bigskip

\noindent
We consider, in particular, a $^{23}\mathrm{Na}+^{39}\mathrm{K}$-mixture [label ``$a$" (``$b$") being associated to $^{23}\mathrm{Na}$ ($^{39}\mathrm{K}$)], featuring $a_a\approx 52\,a_0$ and $a_b\approx 7.6 \, a_0$ (see the caption of Fig. \ref{fig:Trajectory} for the other microscopic parameters). This specific mixture is particularly suitable to test our analytical model because $m_b>m_a$ and $a_b<a_a$, meaning that the massive cores are relatively small and heavy thus constituting a good physical implementation of a ``massive point vortex".  We present the results of our numerical simulations, together with some observables and indicators extracted upon post-processing the raw numerical data as suitable. 

We employ imaginary time propagation [i.e. Eq. (\ref{eq:H_GPE}) following the substitution $t\to-i\tau$] to prepare a state where $a$-component, confined in a disk-like domain of radius $R$, features an off-centered vortex, while $b$-component is localized within its core. This state is initially generated by imprinting a non-trivial phase field (unitary winding number) in $\psi_a$, and is stabilized by the immiscible character ($g_{ab}>\sqrt{g_a g_b}$) of the two BECs. To avoid undesired drifts of the vortex-massive-core complex towards the trap boundary \cite{Guilleumas2001} (we recall that the imaginary-time propagation is a fictitious dynamics and does not conserve the energy), we make use of a strong and sharp Gaussian pinning potential acting only on $\psi_a$ and whose center coincides with the center of the $b$-component wavepacket. We also consider a frame rotating at angular frequency $\Omega$, a circumstance corresponding to the introduction of the term $-\Omega \hat{L}_z$ (where $\hat{L}_z$ is the third component of the angular-momentum operator) in Eq. (\ref{eq:H_GPE}) for both components. The value of $\Omega$ is chosen according to the massive point vortex model (\ref{eq:EL_vortices}) and (\ref{eq:EL_cores}) to support a non-zero precession velocity $\dot{\bm{r}}_{c,v}=\bm{\Omega} \times \bm{r}_{c,v}$ of the vortex-massive-core complex.

The wavefunctions output by the imaginary-time propagation are taken as initial conditions for the real-time evolution, which is performed in the absence of both the pinning potential and the rotation frequency. We remark that, in order to excite the system's normal modes (see also Sec. \ref{sub:Numerical_trajectories} and Sec. \ref{sub:Numerical_normal_modes}), $\psi_b$ is preliminary perturbed as follows: $\psi_b\to\psi_b^\prime:=\psi_b e^{i\bm{r}\cdot \bm{\alpha}}$, where $\bm{\alpha}=m_b(\Delta\dot{\bm{r}}_c)/\hbar$ and $\Delta\dot{\bm{r}}_c$ is the perturbation which is imparted to the core's velocity. In the absence of this perturbation, i.e. for $\bm{\alpha}=\bm{0}$, the vortex and its massive core would exhibit a simple uniform precession around the origin.

\subsection{Trajectories}
\label{sub:Numerical_trajectories}
While performing the real-time evolution of $\psi_a$ and of (the perturbed) $\psi_b$, we track and record the positions $\bm{r}_c(t)$ and $\bm{r}_v(t)$ of both the massive core and its hosting vortex. The former is computed according to the well-known formula
\begin{equation}
\label{eq:Numerical_r_c}
    \bm{r}_c=\frac{1}{N_b}\int \bm{r}|\psi_b|^2\,\mathrm{d}^2\bm{r}
\end{equation}
while the latter demands a different approach since the vortex corresponds to a density \emph{depletion}. One may define $\bm{r}_v$ as the position of the local minimum of $|\psi_a|^2$, but this choice would lead to a rather noisy output, reminiscent of the discrete sampling of $\psi_a$ (i.e. of the real-space mesh). Instead, we found it convenient to follow the following procedure: \textit{i)}: construct $\tilde{\psi}_a$ by cropping $|\psi_a|^2$ to a circle of radius $0.9\,R$ thus getting rid of the tiny annulus (having width $\sim\xi_a$, where $\xi_a$ is the healing length associated to $\psi_a$) over which the density profile $|\psi_a|^2$ decays from its bulk value down to zero; \textit{ii)} Determine $\mathcal{M}_a:=\max|\tilde{\psi}_a|^2$ and define the new density-like distribution $|\Psi_a|^2:=\mathcal{M}_a-|\psi_a|^2$, which is positive definite by construction. In this way, density depletions in $|\psi_a|^2$ are mapped to density peaks in $|\Psi_a|^2$ thus naturally offering the possibility of using a formula of the type (\ref{eq:Numerical_r_c}); \textit{iii)} The vortex center can now be defined as 
\begin{equation}
\label{eq:Numerical_r_v}
    \bm{r}_v=\frac{\int \bm{r}|\Psi_a|^2\,\mathrm{d}^2\bm{r}}{\int |\Psi_a|^2\,\mathrm{d}^2\bm{r}},
\end{equation}
which manifestly involves the \emph{weighted average} of a fictitious density distribution. 

\begin{figure}[h!]
    \centering
    \includegraphics[width=\columnwidth]{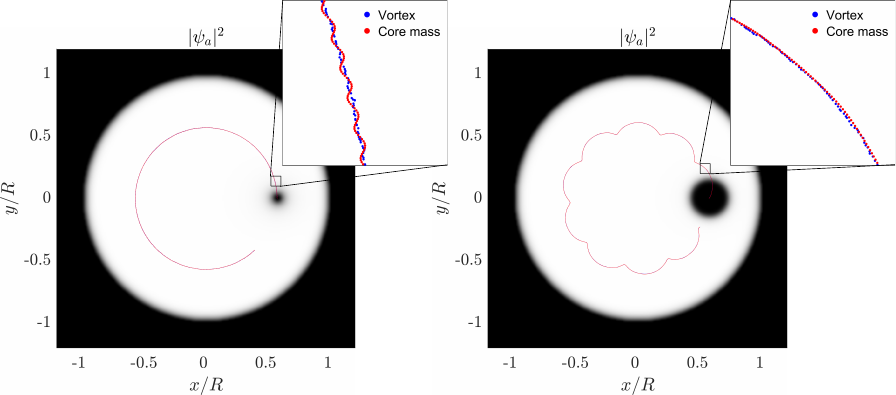}
\caption{Trajectories of the quantum vortex and its massive core as extracted from numerical simulations of coupled GP equations (\ref{eq:H_GPE}) according to formulas (\ref{eq:Numerical_r_c}) and (\ref{eq:Numerical_r_v}). Left panel: light-core regime, obtained for $N_b=1$. The inset shows that the massive core oscillates within its quantum vortex, as predicted by the analytical model [see Fig. \ref{fig:Hybrid_normal_modes} (b) and Sec. \ref{sub:Light_core}]. Right panel: moderate-weight-core regime, obtained for $N_b=5\times 10^3$. $N_a=10^5$.
In agreement with the analytical model, the trajectories feature overall radial oscillations resulting in a multi-lobed trajectory [see Fig. \ref{fig:Hybrid_normal_modes} (d) and Sec. \ref{sub:Eigenfrequencies_off_centered_vortex}]. But, as opposed to the predictions of the analytical model, there is no trace of the rapid precession of the quantum vortex around its massive core (see inset). The microscopic model parameters are listed in the caption of Fig. \ref{fig:Trajectory}.}
    \label{fig:GPE_trajectories}
\end{figure}

The plot of $\bm{r}_c(t)$ and $\bm{r}_v(t)$ as extracted from the real-time simulations of Eq. (\ref{eq:H_GPE}) according to formulas (\ref{eq:Numerical_r_c}) and (\ref{eq:Numerical_r_v}) is illustrated in Fig. \ref{fig:GPE_trajectories} for two different values of $N_b$.
For small values of $N_b\ll N_a$ (see left panel), the massive core oscillates within its hosting quantum vortex, which, instead, travels along an (almost) unaffected uniform circular orbit. This phenomenology is qualitatively analogous to that illustrated in Fig. \ref{fig:Hybrid_normal_modes} (b) and discussed in Sec. \ref{sub:Light_core}. 

For larger values of $N_b/N_a$, the analytical model predicts radial oscillations of the massive-core-quantum-vortex complex associated to a very rapid precession of the quantum vortex around its massive core [see Fig. \ref{fig:Hybrid_normal_modes} (d)]. Our extensive numerical experiments indeed confirm the occurrence of the former phenomenon, resulting in a multi-lobed overall trajectory, but does not provide evidence of the latter (see right panel of Fig. \ref{fig:GPE_trajectories}). We argue that this discrepancy is not due to the thickness of the adopted spatial mesh, which should indeed allow to resolve this rapid precession, if it was present. Instead, we believe that the total lack of this feature is intrinsic to the physical system under investigation and maybe ascribed either to a fast dissipation of such oscillations into sound, or to the partial validity of the assumptions of the analytical model presented in Sec. \ref{sec:Model}, i.e. the fact that the vortex profile is actually not static, but can deform so to adjust, at any time, to the instantaneous effective potential landscape represented by the peaked $|\psi_b|^2$.  

Altogether, the trajectories illustrated in Fig. \ref{fig:GPE_trajectories} disclose rather elusive phenomena, either because the relative motion between the vortex and its massive core is not visible (moderate-mass-core case, see right panel of Fig. \ref{fig:GPE_trajectories}) in GP simulations, or because such relative displacement may be so small (we recall that it would be of the order of the $a$-component healing length) that a direct experimental detection may be not possible (light-mass-core case, see left panel of Fig. \ref{fig:GPE_trajectories}). As we will show in Sec. \ref{sub:Numerical_normal_modes}, this is not the whole story, since there exists a well-defined and experimentally accessible physical observable which unambiguously discloses the signature of the \emph{finite} coupling between the vortex and its massive core.

\subsection{Lobes frequency}
\label{sub:Numerical_normal_modes}
As already mentioned in Sec. \ref{sub:Eigenfrequencies_off_centered_vortex} (see also the right panel of Fig. \ref{fig:Eigenfrequencies_vs_gab}), in the moderate-mass-core case, the eigenfrequency $\omega_2$, which is proportional to the number of lobes in the overall trajectory of the quantum-vortex-massive-core complex, strongly depends on the intercomponent repulsion $g_{ab}$. We remark that this non-trivial dependence could \emph{not} be appreciated in the framework of the previously proposed massive-point-vortex model \cite{Richaud2020,Richaud2021,Richaud2022r_k} as $g_{ab}$ did not play any explicit role (one only relied the assumption that $g_{ab}$ was large enough to ensure a tight confinement of the massive core within its hosting quantum vortex).

In agreement with the analytical model presented in Sec. \ref{sec:Model} and \ref{sec:Normal_modes}, our numerical simulations indeed show that $\omega_2$ increases upon increasing $g_{ab}$ and eventually saturates to a limiting value. 

To show this, we analyze the raw data $r_c(t):=\sqrt{x_c(t)^2+y_c(t)^2}$ extracted from real-time GP simulations for different values of $g_{ab}$, but for the same value of $N_b$ and of the initial perturbation $\bm{\alpha}$. 
\begin{figure}[h!]
    \centering
    \includegraphics[width=\columnwidth]{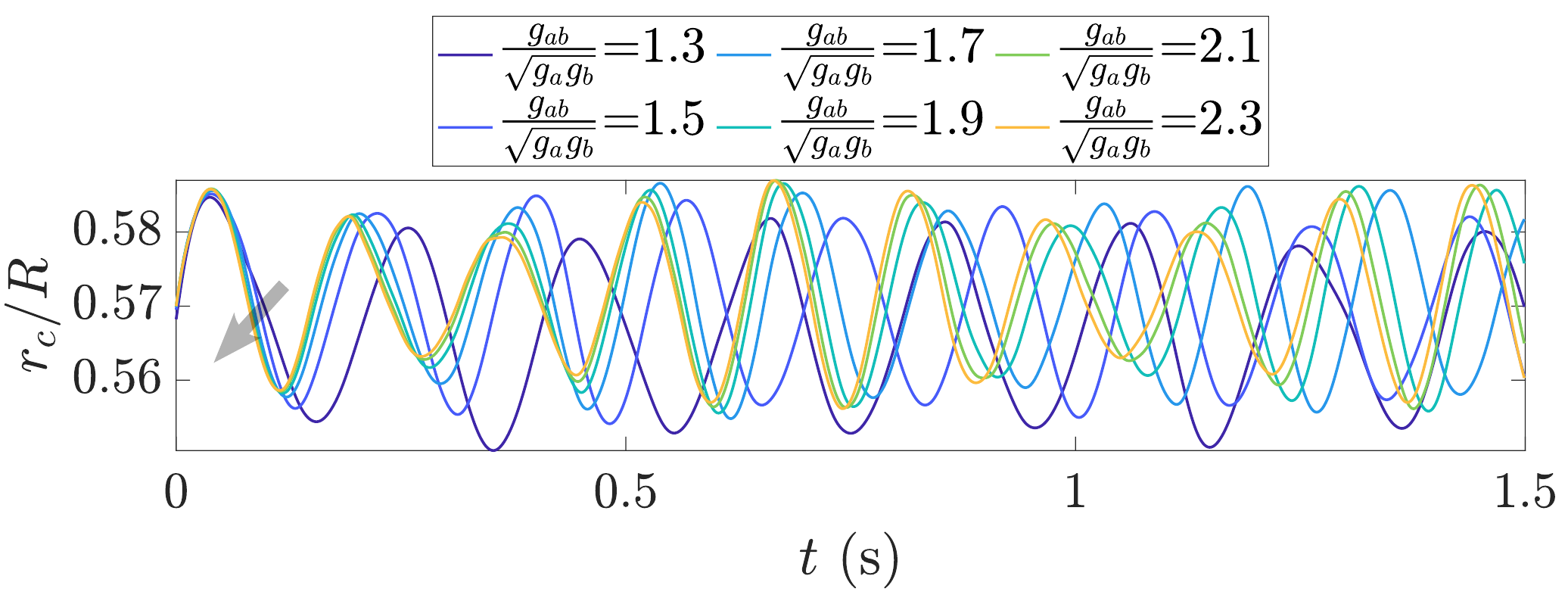}
    \caption{Radial distance of the massive core as function of time for different values of $g_{ab}$. Each peak of the function $r_c(t)$ corresponds to a lobe of a trajectory of the type illustrated in the right panel of Fig. \ref{fig:GPE_trajectories}. As pictorially illustrated by the gray arrow, increasing $g_{ab}$, the period of the sine-like waves gets smaller, that means that its inverse, i.e. $\omega_2$, gets larger. The main frequency of each sine-like wave is extracted by FFT and illustrated in Fig. \ref{fig:omega_2_vs_gab} as a function of $g_{ab}$. The assumed microscopic parameters are listed in the caption of Fig. \ref{fig:Trajectory} and at the beginning of Sec. \ref{sec:GP_results}, with $N_a = 9\times 10^4$ and $N_b =2\times10^3$.}
    \label{fig:r_vs_t_sweep_gab}
\end{figure}
While the rapid increase of $\omega_2$ with $g_{ab}$ is already evident from the observation of Fig. \ref{fig:r_vs_t_sweep_gab}, a more quantitative observation can be made by applying the Fast Fourier Transform (FFT) to $r_c(t)$, which ultimately constitutes the discrete sampling of a continuous signal. The output of (the absolute value of) this transform is a distribution peaked around a certain frequency, which is identified with $\omega_2$ and illustrated in Fig. \ref{fig:omega_2_vs_gab} as a function of $g_{ab}$.  

\begin{figure}[h!]
    \centering
    \includegraphics[width=0.5\columnwidth]{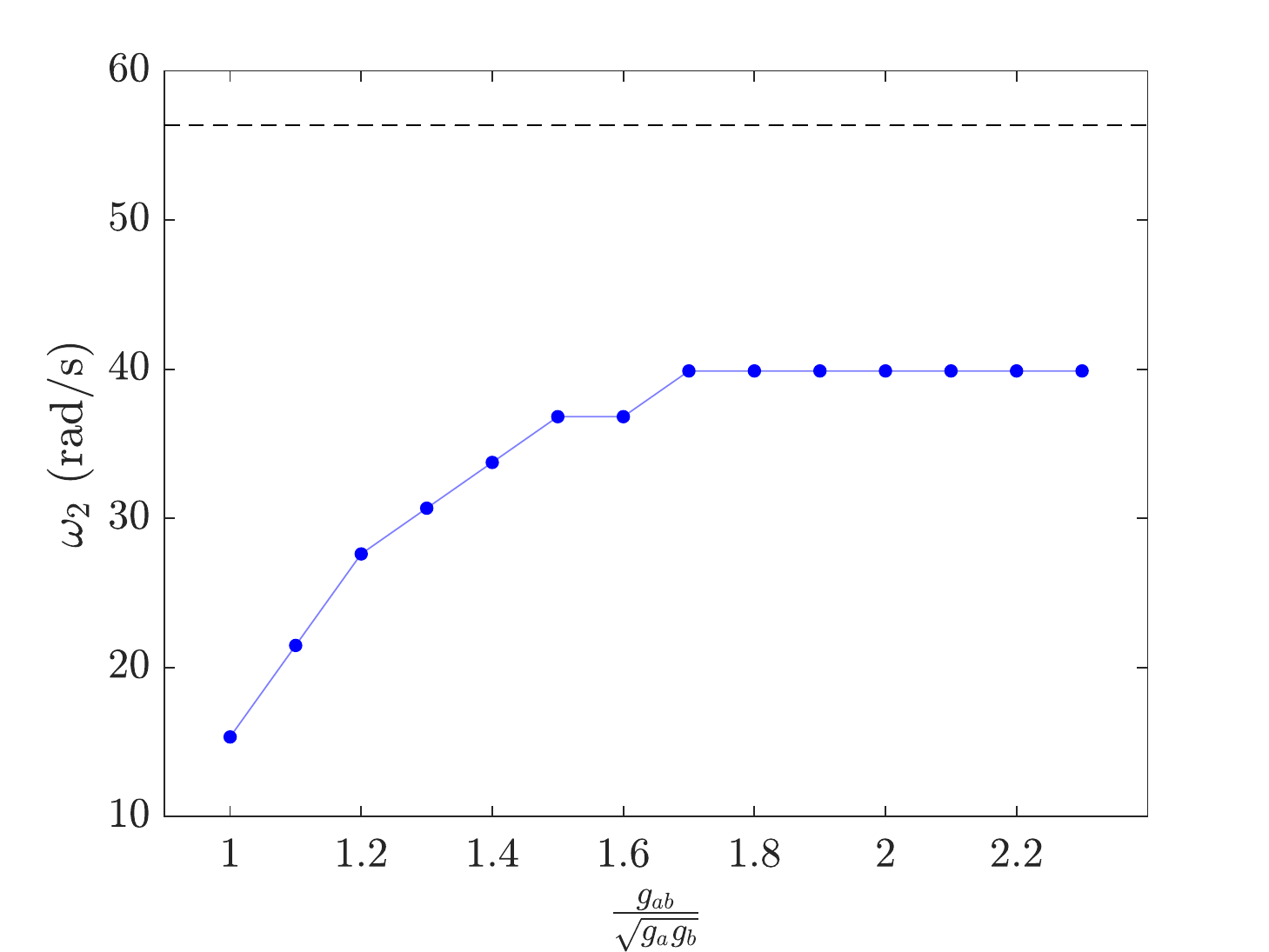}
    \caption{Functional dependence of the numerically-determined frequency of the radial oscillations of the quantum-vortex-massive-core complex on the interspecies interaction $g_{ab}$. The value of $\omega_2$ corresponds to the frequency at which the (absolute value of the) FFT of $r_c(t)$ (see Fig. \ref{fig:r_vs_t_sweep_gab}) is peaked. The gray dashed line corresponds to the $g_{ab}$-independent relation (\ref{eq:omega_2_*}) i.e. to the model \cite{Richaud2021} where the massive core was assumed to be always centred at the centre of the quantum vortex. The assumed microscopic parameters are listed in the caption of Fig. \ref{fig:Trajectory} and at the beginning of Sec. \ref{sec:GP_results}.}
    \label{fig:omega_2_vs_gab}
\end{figure}

For comparison, in the aforementioned figure, we also illustrate the $g_{ab}$-independent value corresponding to Eq. (\ref{eq:omega_2_*}),  derived in Ref. \cite{Richaud2021} within a massive-point-vortex model which did not account for the possible relative motion between the vortex and its massive core. While this quantity was shown to be the asymptotic value which $\omega_2$ tends to for $g_{ab}\to+\infty$ according to the analytical model discussed in Sec. \ref{sec:Model} and \ref{sec:Normal_modes} (see also the right panel of Fig. \ref{fig:Eigenfrequencies_vs_gab}), our GP simulations show that the numerically-computed values of $\omega_2$ actually tend to a smaller limiting value for large values of $g_{ab}$. This quantitative, but not qualitative, discrepancy with respect to the analytical predictions is somehow intrinsic to a variational point-vortex model, as the vortex with filled massive core is not really point-like in GP simulations, but has a finite typical size. We indeed verified that, upon increasing the microscopic parameters $g_{ab}$ and $g_a$, the typical size of vortices in $\psi_a$ gets increasingly smaller and that the asymptotic value which the relevant relation $\omega_2(g_{ab})$ tends to, is quantitatively well captured by Eq. (\ref{eq:omega_2_*}).

\begin{figure}[h!]
    \centering
    \includegraphics[width=0.7\columnwidth]{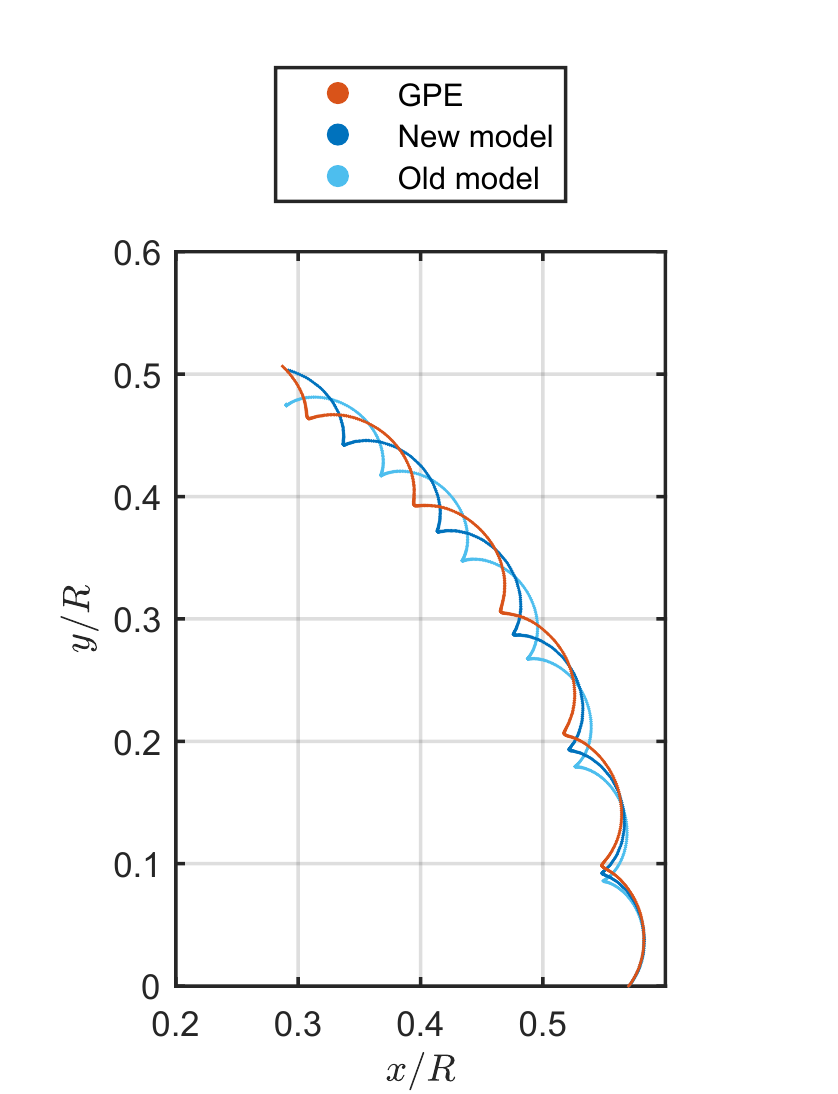}
    \caption{Comparison of the GPE solution with the previous and the here extended point-like model. Our new model improves the prediction of the macroscopic radial oscillations frequency. The GPE core-trajectory and our model prediction show approximately five complete oscillations and a half, wheras the old model predicts six complete oscillations.
    For illustrative purposes we choose a some higher $g_{ab}$ and $g_a$ values, with $g_{ab}/g_a$ the same as at the beginning of Sec. \ref{sec:GP_results}; this way, we select a naturally more ``point-like" regime. Here: $N_a = 10^5$,
    $N_b = 2\times 10^3$.}
    \label{fig:comp_of_models}
\end{figure}

Figure \ref{fig:comp_of_models} shows a result of a GPE simulation and our model prediction, as compared to the previous model's one. As shown, our model leads to a lobe frequency that is closer to the GPE solution; 
for illustrative purposes we choose a some higher $g_{ab}$ and $g_a$ values, with the same $g_{ab}/g_a$ as at the beginning of Sec. \ref{sec:GP_results}; this way, we select a naturally more ``point-like" regime.
   
\section{Concluding Remarks}
\label{sec:Conclusions}
Motivated by recent theoretical results \cite{Richaud2020,Griffin2020,Richaud2021,Richaud2022r_k,Caldara2023} according to which the dynamics of quantum vortices with filled massive cores requires second-order acceleration terms ensuing from the inertial character of the core-filling component, we studied the impact of a \emph{finite} coupling energy between the vortex and its massive core. In fact, the basic assumption which previous models \cite{Richaud2020,Griffin2020,Richaud2021,Richaud2022r_k,Caldara2023} relied on, i.e. that the center of the massive core always coincides with that of the hosting quantum vortex, may seem a priori questionable, especially when the immiscibility condition $g_{ab}>\sqrt{g_a g_b}$ is barely met.

We thus develop a point-vortex-model where the quantum vortex and its massive core are described by different sets of dynamical variables. It turns out that the quantum vortex is still described by first-order motion equations, as it is customarily in superfluid vortex dynamics \cite{Fetter1998,Kim2004}, but it is harmonically coupled to a massive particle, which is, in turn, described by Newton's second law. The effective point-like motion equations for the composite object are extracted from the two coupled GP equations according to a standard time-dependent variational Lagrangian approximation scheme \cite{PerezGarcia1996,Kim2004}, where the vortex is modelled by a Gaussian-like density depletion from an otherwise uniform density distribution, and the massive core by a Gaussian wavepacket. The effective ``spring constant" associated to the vortex-massive-core interaction is shown to depend on a number of microscopic parameters, including the interspecies repulsion and the typical widths of the aforementioned Gaussian-like density distributions.  

While the predictions of the obtained dynamical model correctly reproduce those of the more simple massive-point-vortex-model (see Refs. \cite{Richaud2020,Richaud2021,Richaud2022r_k}) in the limit $g_{ab}\to +\infty$, some phenomena that we observed in GP simulations can be correctly explained only within this more refined framework. One of them, which is also amenable to experimental detection with current technology, concerns the frequency at which the quantum-vortex-massive-core complex oscillates when it is displaced radially from its equilibrium orbit \cite{Richaud2021}: it steadily increases upon increasing the stiffness of the effective spring, before eventually saturating to a constant value when the effective spring gets so stiff not to allow any relative displacement.    

The developed analysis is expected to provide a reliable effective dynamical model for the dynamics of vortices with filled massive cores, especially in those regimes where the immiscibility condition is barely met. This is possibly the case of thermal atoms trapped within the core of a vortex in a dilute atomic BEC at finite temperature \cite{Coddington2004} (see also chapter 9 of Ref. \cite{Griffin2009}), or of the quasi-particle bound states of a vortex in a fermionic superfluid \cite{Kwon2021}. It should also stimulate the already very active research in the real-time dynamics of few-vortex systems \cite{Serafini2017,Kwon2021} and in two-component vortices \cite{Law2010,Gallemi2018,Wang2018,Ruban2021,Ruban2022,Ruban2022PRA,Choudhury2022,Zhu2022,Wang2022,Kasamatsu2003,Katsimiga2023}.

Possible future research directions include the extension of the current model to vortex lines in 3D superfluid samples \cite{Schwarz1985,Galantucci2020}, as well as the adoption of more sophisticated ansatzes in the derivation of the analytical point-vortex model (the very recent Ref. \cite{Doran2022} suggested, for example, that the vortices' and the cores' profiles are better described by super-Gaussian wavefunctions for large atom numbers of the in-filling component, a circumstance which may possibly result in an \emph{anharmonic} effective coupling). Also, one may want to excite some internal modes of the in-filling component (e.g. the monopole and the quadrupole mode \cite{PerezGarcia1996,Stringari1996}) and investigate their possible effect on the overall motion of the vortex-massive-core complex. Eventually, the proposed analytical point-vortex model may be further generalized to account for possible dissipative effects \cite{Billam2015} or to systems featuring an additional inter-component coherent coupling \cite{Choudhury2022}.

\section*{Acknowledgements}
A. R. received funding from the European Union’s
Horizon research and innovation programme under the Marie Skłodowska-Curie grant agreement \textit{Vortexons} no. 101062887. Computational resources were provided by HPC@POLITO (http://hpc.polito.it).

\begin{appendices}

\section{Inclusion of the spring-like energy term in the point-like model}
\label{EabComputation}


We perform the derivation of the $g_{ab}$-dependent energy term describing the vortex-mass coupling in Lagrangian (\ref{eq:L_tot}).
The ansatz is (see also equations (\ref{eq:psi_a})-(\ref{eq:psi_b}))

\begin{equation}
    \psi_a(\bm{r},t)=\sqrt{n_a-n_a e^{-|\bm{r}-\bm{r}_v(t)|^2/\sigma_a^2 }
    }\,e^{i\theta_a}
\end{equation}
\begin{equation}
    \psi_b(\bm{r},t) = \left( \frac{N_b}{\pi\sigma_b^2} \right)^{1/2}
    e^{-\frac{|\bm{r}-\bm{r}_{c}(t)|^2}{2\sigma_b^2}} e^{i\bm{r}\bm{\alpha}_i(t)},
\end{equation}
for $|\bm{r}|<R$, with
\begin{equation}
    \theta_a= \arctan\left(\frac{y-y_v}{x-x_v}\right)-\arctan\left(\frac{y-y_v^\prime}{x-x_v^\prime}\right).
\end{equation}

The corresponding densities satify the conditions
\begin{equation}
    \int d^2r |\psi_a|^2=N_a, 
    \quad  \int d^2r |\psi_b|^2=N_b.
\end{equation}

With the narrow-Gaussian approximation, the integration domain becomes $[-\infty,+\infty]$. After substituting the ansatz, it follows:

\begin{equation}
    E_{ab}=g_{ab}\left[n_a N_b- n_a N_b \sigma_{ab} e^{-\frac{|\bm{r}_{c}(t)-\bm{r}_{v}(t)|^2}{\sigma_a^2+\sigma_b^2}}\right],
\end{equation}
with $\sigma_{ab}=\frac{\sigma_a^2}{\sigma_a^2+\sigma_b^2}$.
For the dynamics evaluation purpose, the constants can be neglected. Also, the exponential can be approximated by a first-order Taylor expansion, since the relative variation of the vortex and core mass centres is assumed very small due to immiscibility. Thus:
\begin{equation}
    E_{ab}=g_{ab}n_a N_b\sigma_{ab}|\bm{r}_c(t)-\bm{r}_v(t)|^2
\end{equation}
in other words, the coupling energy is comparable to a spring coupling the two species, which constant depends on the system's parameters. 
Finally, note that the higher $g_{ab}$ or $\sigma_{ab}$ (i.e. the inter-species repulsion), the stiffer the spring, or the more rigid the core confinement.
\\This computation can be easily extended to the case of more vortices. Here, a summation over the number of vortices appears in $\psi_a$ and $\psi_b$.
Again, with the narrow-Gaussian approximation, all the multiple summations reduce to a single-index summation and the integration domain becomes $[-\infty,+\infty]$.


\section{Circular-orbit solutions of Lagrange equations}
\label{app1}

Equations (\ref{eq:EL_vortices}) and (\ref{eq:EL_cores}) exhibit constant-radius solutions that depend on the angular velocity $\Omega$ and on the other model parameters. By substituting expressions (\ref{C1}) and (\ref{C2}) in Eqs. (\ref{eq:EL_vortices}) and (\ref{eq:EL_cores}), one obtains the equations for the radii ${\tilde r}_c$ and ${\tilde r}_v$
\begin{equation}
m_b N_b \Omega^2 {\tilde r}_c = 
2K ({\tilde r}_c -{\tilde r}_v), \quad K = g_{ab} \sigma_{ab} n_a N_b,
\label{circ1}
\end{equation}
\begin{equation}
n_a h \Omega {\tilde r}_v =
\frac{h^2 n_a {\tilde r}_v }{4m_a\pi(R^2-{\tilde r}_v^2)} +
2K ({\tilde r}_c-{\tilde r}_v),
\label{circ2}
\end{equation}
which one easily recasts into the form
$$
{\tilde r}_c = \frac{ 2K}{2K - m_b N_b \Omega^2 } {\tilde r}_v, 
$$
$$
\left [ ( n_a h \Omega +2K ) - \frac{h^2 n_a/m_a }{4\pi(R^2-{\tilde r}_v^2)} 
-\frac{ 4K^2 }{2K - m_b N_b\Omega^2 } \right ] {\tilde r}_v =0.
$$
For $\Omega = 0$, the first equation
gives ${\tilde r}_c = {\tilde r}_v$
while the second entails ${\tilde r}_v=0$: both the mass and the vortex are placed at the origin and no motion takes place. For $g_{ab} \to \infty$
one recovers the model where, the vortex is integral with the mass (${\tilde r}_v \to {\tilde r}_c$) due to the infinite strength $K$ of the harmonic force. The characteristic frequencies (\ref{eq:Omega_plus}) and (\ref{eq:Omega_minus}) of the circular orbits for this limiting model are easily obtained from the second equation.

\section{Hamilton equations}
\label{app2}

The dynamical equations relevant to Hamiltonian (\ref{eq:H}) can be easily found by means of Poisson brackets (\ref{DB}). The latter preserve their structure when introducing the canonical coordinates $X_c$, $Y_c$, $P_X$, $P_Y$, $X_v$, and $Y_v$ of the rotating frame [see equation (\ref{rot})]
where Hamiltonian $H$ takes the form 
$$
H =
\frac{ P^2_X + P^2_Y  }{2M_b} - \Omega ( X_c P_Y -Y_c P_X) 
+
\frac{h n_a}{2}  \Omega ( X_v^2 + Y_v^2) 
$$
$$
+
K \Bigl [ (X_v -X_c)^2 +(Y_v-Y_c)^2)  \Bigr ]
+
\frac{ n_a h^2 }{4\pi m_a} {\rm ln} \left [ 
\frac{ R^2 -X_v^2-Y_v^2   }{ R^2}  \right ]
$$
with $K = g_{ab} \sigma_{ab} n_a N_b$ and $M_b=m_b N_b$. Then the explicit expression for the Hamilton equations ${\dot {\bf Z}} = {\mathbb E} (\nabla H)^T$, discussed in section (\ref{LINR}), is
$$
{\dot X}_v = \frac{2K}{n_a h} (Y_v- Y_c) + \Omega Y_v
- \frac{h }{2\pi m_a} \frac{ Y_v }{R^2 -X_v^2-Y_v^2},
$$
$$
{\dot Y}_v = - \frac{2K}{n_a h} (X_v- X_c) - \Omega X_v
+ \frac{h }{2\pi m_a} \frac{ X_v }{R^2 -X_v^2-Y_v^2}.
$$
$$
{\dot X}_c = \frac{P_X}{ M_b } +\Omega Y_c,
\qquad
{\dot Y}_c = \frac{P_Y}{ M_b } -\Omega X_c,
$$
$$
{\dot P}_X = 2 K (X_v -X_c) +\Omega P_Y,
$$
$$
{\dot P}_Y = 2K (Y_v -Y_c) -\Omega P_X.
$$
The fixed-point solutions of these equations 
$$
{\bar X}_v, \, {\bar Y}_v, \,
{\bar X}_c, \, {\bar Y}_c, \, {\bar P}_X, \, {\bar P}_Y,
$$
reproduce 
formulas (\ref{circ1}) and (\ref{circ2}), with
${\tilde r}^2_c = {\bar X}_c^2+{\bar Y}_c^2$ and ${\tilde r}^2_v = {\bar X}_v^2+{\bar Y}_v^2$, and thus identify with constant-radius solutions. 

Perturbed solutions representing deviations from fixed points are found by introducing new local variables such that
$$
X_c = {\tilde X}_c + \xi_x, \quad Y_c = {\tilde Y}_c + \xi_y ,
$$
$$
P_X = {\tilde P}_X + \pi_x, \quad P_Y = {\tilde P}_Y + \pi_y,
$$
$$
X_v = {\tilde X}_v + \eta_x, \quad
Y_c = {\tilde Y}_c + \eta_y,
$$
In the new picture involving perturbation variables $\xi_x$, $\xi_y$, $\pi_x$, $\pi_y$, $\eta_x$ and $\eta_y$ it is advantageous to write Hamiltonian $H$, up to a constant term $H_0$ in the quadratic form
$$
H_2 = H_0
+  A_{11}\eta_x^2  + A_{22} \eta_y^2 + 2A_{12} \eta_x \eta_y
$$
$$
+
2A_{13} \eta_x \xi_x  + 2A_{24}  \eta_y \xi_y 
+
A_{33}  \xi_x^2  + A_{44} \xi_y^2   
$$
$$
A_{55} \pi_x^2 + A_{66} \pi_y^2 
+ 2A_{36}  \xi_x  \pi_y +  2 A_{45}  \xi_y \pi_x
$$
where the elements $A_{mn}$ ($m$, $n \in [1,6]$) of the symmetric Hessian matrix ${\mathbb H}= A$ are given by
$$
A_{11} =  K + \frac{ h n_a }{2 } \Omega + 
\frac{h^2 n_a }{4\pi m_a}  \frac{ {\tilde Y}_v^2 - {\tilde X}_v^2 -R^2}{(R^2 -{\tilde r}_v^2)^2},
$$
$$
A_{22} =  K + \frac{ h n_a }{2 } \Omega +
\frac{h^2 n_a }{4\pi m_a}   \frac{ {\tilde X}_v^2 - {\tilde Y}_v^2 -R^2}{(R^2 -{\tilde r}_v^2)^2},
$$
$$
A_{12}  =
- \frac{h^2 n_a }{4\pi m_a}  \frac{ 2{\tilde X}_v {\tilde Y}_v }{(R^2 -{\tilde r}_v^2)^2},
$$
$$
A_{33} = A_{44} =  K, \quad 
A_{55} = A_{66} =  \frac{1}{2M_b}, \quad
$$
$$
A_{13} = A_{24} =  -K, \quad 
A_{36} =  -\frac{ \Omega }{2 } , \quad A_{45} =  \frac{ \Omega }{2 }.
$$

The elements not included in this list are zero. The motion equations
for the perturbative solutions then read 
$$
{\dot {\bf V}} = {\mathbb E} (\nabla H_2)^T
$$
where ${{\bf V}} = (\eta_x, \eta_y, \xi_x, \xi_y, \pi_x, \pi_y)^T$
and
$$
(\nabla H_2)^T= 2
{\left [ \begin{array}{cccccc}
A_{11}  & A_{12} & A_{13} & 0 & 0 & 0 \\
A_{21}  & A_{22} & 0 & A_{24}& 0 & 0  \\
A_{31}  & 0 & A_{33} & 0 & 0 & A_{36} \\
0  & A_{42} & 0 & A_{44}& A_{45} & 0   \\
0  & 0 & 0 & A_{54}& A_{55} & 0  \\
0  & 0 & A_{63} & 0 & 0 & A_{66} \\
\end{array} \right] }
\, \left [ \begin{array}{c}
\eta_x \\
\eta_y \\
\xi_x \\
\xi_y \\
\pi_x \\
\pi_y \\
\end{array} \right] 
$$
The stability character and other properties of the ensuing perturbative solutions are discussed by computing the eigenvalues of the matrix associated to this linear dynamical system.

\end{appendices}


\end{document}